\documentclass[preprint]{aastex}
\begin{document}
\title{Kinematic Projecting of Pulsar Profiles}
\author{V.A.~Bordovitsyn}
\affil{Tomsk State University, Tomsk Russia}
\author{ V.Ya.~Epp}
\and
\author{V.G.~Bulenok}
\affil{Tomsk State Pedagogical University, Tomsk, Russia}

\begin{abstract}

We suggest a technique for
construction of profiles of incoherent pulsar emission using
instantaneous angular distribution indicatrix of radiation
from relativistic source moving in the neutron star magnetosphere. In
the general case the indicatrix depends on the kinematic parameters of
radiation source such as velocity and acceleration.
The method is illustrated by calculation of the profiles of pulsar
radiation based on synchrotron radiation indicatrix. This technique
can be easily generalized for the case of  other types of
relativistic particles radiation. The considered profiles are compared
with observed ones. A good agreement was found with profiles of
$\gamma$-pulsars --- Geminga, Crab and Vela.

\end{abstract}

\section{Introduction}

There is no clear conception of the nature
of pulsar elec\-tro\-mag\-ne\-tic radiation. We accept the
well known point of view suggested  by I.S.~Shklovsky~\cite{1}
that at least the high
frequency part of radiation represents the incoherent
synchrotron radiation.

There are different models of pulsar radiation based on
various physical processes which are sometimes
contradictory and insufficiently conditioned.  We suggest a
technique of kinematic projecting of synchrotron
radiation~\cite{2} which gives the pulsar radiation profiles including
the phase dependence of polarization characteristics~\cite{3,4,5}.
Then one can analyze the coherence effects and possible
mechanisms of pulsar radiation.

The suggested technique of determination of the pulsar
radiation profiles is as a matter of fact a kinematic projecting of
relativistic radiation regardless to the specific source of
the radiation.  The role of such sources of radiation can play
relativistic charged particles electrons, magnetic particles (neutrons)
or even bunches of those particles. The main source of information in
this technique is the highly directed indicatrix of the angular
distribution of instantaneous power of relativistic radiation, which is
specified by the kinematic and dynamic characteristics of the emitting
object in every specific case.
It is known~\cite{1,6} that there are two models of radiation
geometry.  The first one is due to two beams of curvature
radiation associated with the magnetic poles.  The second one is
associated with synchrotron radiation in a plane orthogonal to the
magnetic axis of pulsar.  We restrict our consideration with the
second model of radiation, though it is possible to apply this
technique in the first case too.

\section{Transformation of angular distribution into the pulse profile}

We start from the indicatrix of synchrotron radiation  averaged over
the period. Because of high energy main part of radiation is
concentrated near the plane of symmetry which we call the light plane.
Directions, assigned to radiation maximum, are situated in the
light plane.  Axis orthogonal to the light plane
coincides with the direction of pulsar magnetic moment $\mu$.  Vector
$\mu$ is precessing around the pulsar axis of rotation $M$ with
the angular velocity $\Omega$.

We assume that the angular velocity $\omega$ of the emitting
particles is much more than $\Omega$.
\begin{equation}
\omega=\frac{eH}{m_0c\gamma}\gg\Omega,
\label{1}
\end{equation}
here $e$ is the charge, $m_0$ is the mass of rest, $c$ is the speed of
light, $\gamma=\sqrt{1-\beta^2}$ is the relativistic Lorentz factor,
$\beta=u/c$, $u$ is the particle velocity. It follows from the last
inequality that
\begin{equation}
\gamma\ll\frac{eH}{m_0c\Omega}.
\label{2}
\end{equation}

Let us estimate the order of $\gamma$-factor satisfying the last
condition. It can be taken for pulsars that $\Omega\approx
10~s^{-1}$, $H\approx 10^{12}$~Oe not far from the surface and
$H\approx 10^{6}$~Oe in the periphery of magnetosphere (near the
light cylinder). Then the condition \ref{2} gives the value $\gamma\ll
10^{12}\div 10^{18}$, i.e.  condition \ref{2} is always
satisfied in available relativistic and even ultrarelativistic case.
\begin{figure}
\begin{center}
\epsscale{0.3}
\plotone{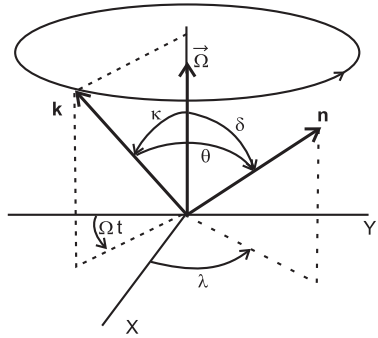}  
\caption{The coordinate system.}
\label{f:1}
\end{center}
\end{figure}

We introduce the coordinate system shown in Fig.~\ref{f:1}.
Direction of the angular velocity of pulsar is parallel to
$Z$-axis.  The pulsar magnetic axis defined by unit vector
${\boldmath k}$ is specified by the angles $\kappa$ and $\Omega t$. The unit
vector of the observer's direction is given by the angles $\delta$ and
$\lambda$. We assume that vector ${\boldmath k}$ is situated in the plane
$YZ$ at the moment $t=0$.

We present the angular distribution of radiation
averaged over the period of particle motion in a form
$$\overline{dW}=W\;\rho(\theta)\;\sin\theta d\theta,$$
where $W$ is the total power of radiation, which is relativistic
invariant, $\rho(\theta)$ is the function of the angular distribution
of radiation normalized to unity
$$\int\limits_0^\pi \rho(\theta)\; \sin\theta d\theta=1.$$

In the case of synchrotron radiation~\cite{3}
$$W=\;\frac{2}{3}\;\frac{e^2}{c}\;\omega^2\beta^2\gamma^4=
W_{\mbox{\scriptsize{SR}}}$$
and
\begin{equation}
\rho(\theta)=\;\frac{3}{8}\;\gamma^{-4}\left\{
\frac{2+\beta^2\sin^2\theta}{(1-\beta\cos\theta)^{5/2}}
-\frac{\sin^2\theta(4+\beta^2\sin^2\theta)}
{4\gamma^2(1-\beta\cos\theta)^{7/2}}\right\}.
\label{0}
\end{equation}

It is obvious that the angle $theta$ varies in the course of time, as the
unit vector ${\boldmath k}$ setting the position of light plane is precessing
around the $Z$ axis. It is easy to show that
\begin{equation}
\cos\theta(t)=\cos\delta\cos\kappa+\sin\delta\sin\kappa\sin(\Omega
t-\lambda).
\label{2'}
\end{equation}
If we know $\theta(t)$, we determine the radiation profile $\rho(t)$ and the
power $\overline{dW}$ emitted in the directions of the angle interval
$\sin\theta d\theta$. The maximum of radiation will be
observed at the moment of intersection of the observation direction
by the light plane when $\cos\theta=0$, i.e. when
\begin{equation}
\sin(\Omega t-\lambda)=-\cot\delta\cot\kappa.
\label{3}
\end{equation}
The latter equation has a solution when
$|\cot\delta\cot\kappa|\leq 1$.

Equation (\ref{3}) has two roots within the
precession period which are equal at $\delta=\pi/2\pm\kappa$.  In other words
the light plane intersects the observation direction twice within the
period, if the angle between the vector ${\boldmath n}$ and plane
$XY$ is less than $\kappa$.

If the light plane intersects the observation direction (or touches
it), then the angle $\lambda$ can be
chosen such that the observation direction has been situated in the
light plane at $t=0$. Under this initial condition  the radiation pulse
takes on the maximum value $\rho(\theta=\pi/2)$ at  $t=0$.  Thereat it
follows from the equation (\ref{3}) that
\begin{equation}
\sin\lambda=\cot\delta\cot\kappa.
\label{4}
\end{equation}
Another form of relation (\ref{4}) is possible
$$\sin\kappa=\frac{\cos\delta}{\sqrt{\cos^2\delta+\sin^2\delta
\sin^2\lambda}}, \quad 0<\kappa,\quad \delta\leq\pi/2.$$

It follows from equation  (\ref{4})that for the same acute angles $\kappa$
and $\delta\leq\pi/2-\kappa$ the term $\sin\lambda$ takes equal
values within the range $0<\lambda<\pi$, i.e.  $\lambda_1$ and
$\lambda_2=\pi-\lambda_1$, which corresponds to two possible
intersections of the observation direction and the light plane. At
$\lambda_1=\lambda_2=\pi/2$ we have $\kappa=\pi/2-\delta$. In this case the
observation direction only touches the light plane. Note that equation
(\ref{4}) is symmetric relative to angles $\kappa$ and $\delta$. It
means that the positions of the two peaks of profiles will not change
if we swap $\kappa$ and $\delta$.
The case of $\kappa=0$ is special. It follows from equation (\ref{2'})
that in this case $\cos\theta=\cos\delta=$const and pulsar is observed
as a source of synchrotron radiation with constant luminosity and
angular distribution of radiation defined by relation (\ref{0}).

In the ultrarelativistic case of $\gamma\gg 1$  the main part of
radiation is concentrated near the light plane
$\theta\approx\pi/2$.  Thereat duration of the radiation pulse is
much less than $\Omega^{-1}$. It allows to expand the function
$\rho(\theta)$ in a series in small angles $\psi=\pi/2-\theta$, or
small $\Omega t$. To a first approximation in $\Omega t$ we get
\begin{equation}
\psi=\cos\delta\cos\kappa+\sin\delta\sin\kappa(\Omega t
\cos\lambda-\sin\lambda).
\label{5}
\end{equation}
Specifying the initial conditions in accordance with equation
(\ref{4}),  we have
\begin{equation}
\psi=\Omega t\sin\delta\sin\kappa\cos\lambda,
\label{6}
\end{equation}
or, with account of (\ref{4}),
$$\psi=\pm\Omega t\sqrt{\sin^2\delta\sin^2\kappa-\cos^2\delta
\cos^2\kappa}.$$
We introduce another angular variable
\begin{equation}
\chi=\gamma\psi=\gamma\overline{\omega}t, \quad
\overline{\omega}=\Omega\sin\delta\sin\kappa\sin\lambda.
\label{7}
\end{equation}
The variable $\overline{\omega}$ has a sense of angular velocity
at which the observer's direction intersects the light plane.
The highest velocity corresponds to passage through the radiation
maximum. In this case $\delta$ and $\lambda$ are associated with the
direction in the light plane.

Substituting equations (\ref{6}) and (\ref{7}) into equation (\ref{0})
we find the radiation profile in a form
\begin{equation}
\rho(t)=\;\frac{3}{32}\;\gamma\;\frac{7+12\chi^2}
{(1+\chi^2)^{7/2}}.
\label{8}
\end{equation}
If $\delta=\pi/2\pm\kappa$ or $\psi=\kappa$ then $\overline{\omega}=0$ and
we have to take into account the terms of
higher order in the expansion in $\Omega t$.

\begin{figure}
\begin{center}
\epsscale{0.8}
\plotone{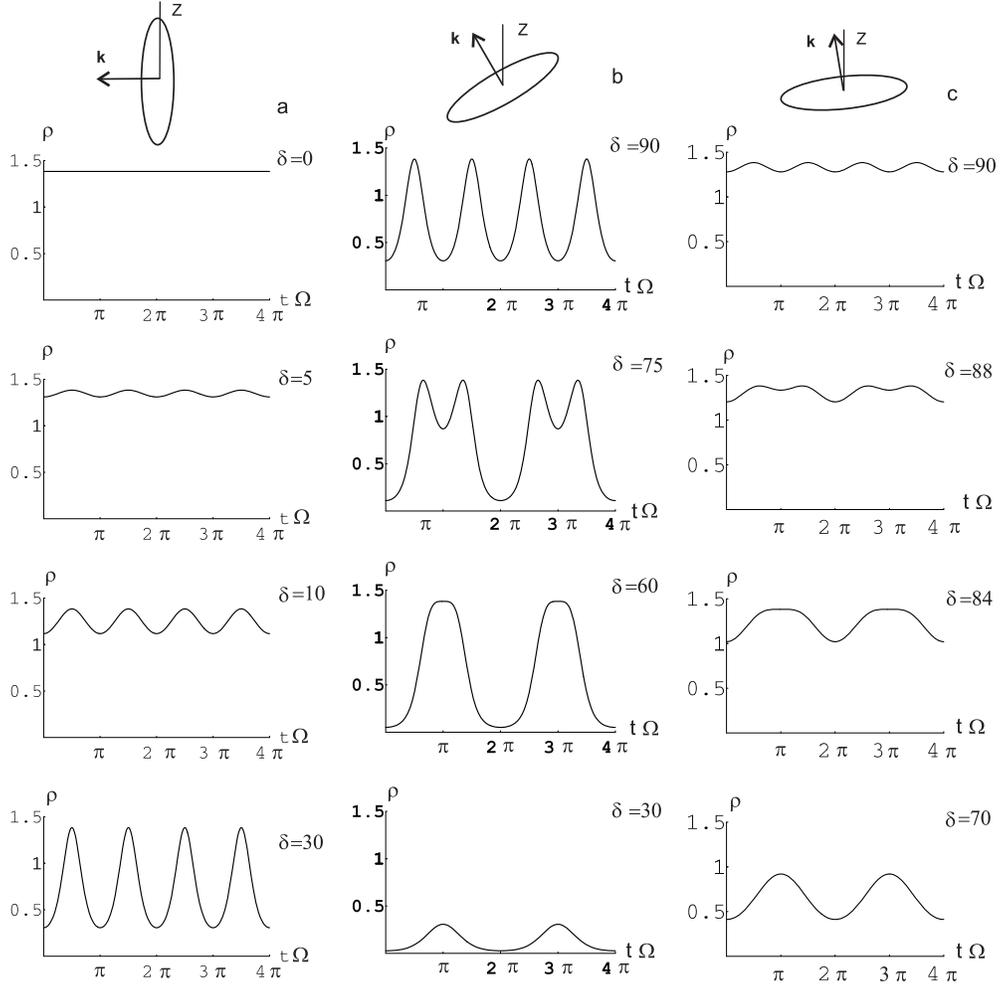}
\caption
{The  radiation phase profiles for $\beta=0.9$
and angles:  a --- $\kappa=90^\circ$; b --- $\kappa=30^\circ$; c ---
$\kappa=6^\circ$}
\label{f:2}
\end{center}
\end{figure}

Figs~\ref{f:2} display the synchrotron radiation profiles correspond to
weakly relativistic electron at $\beta=0.9$ and at various values of
angle $\delta$ and fixed $\kappa$. Fig.~\ref{f:2}~a shows the case
when the angle $\kappa=90^\circ$ i.e. when vector ${\boldmath k}$ is
orthogonal to rotation axis.  If the observer is situated in the
direction of $Z$ axis ($\delta=90^\circ$), radiation of constant
intensity is registered.  As the angle $\delta$ increases the radiation
is modulated at the frequency of $2\Omega$ and the effect of glimmer
pulsar is observed. One can see that the greater is the angle $\delta$
the deeper is the glimmer modulation.
If the angle between pulsar magnetic axis and rotation axis is acute
(see Fig.~\ref{f:2}~b), then the distance between the peaks of radiation
within a period decreases with increasing of angle $\delta$, and at
$\delta+\kappa=\pi/2$ they merge in one maximum. At further
decreasing of $\delta$ the light plane does not intersect the
observer's direction and the amplitude of splashes falls quickly off.
When the angle between the pulsar magnetic axis and rotation
axis is small, the light plane intersects the observation direction
under a small angle (see Fig.~\ref{f:2}~c).
In consequence the splashes duration increases considerably and the depth of
modulation falls off.  Thereby the main part of radiation  is concentrated
near the plane orthogonal to the rotation axis ($\delta\approx
90^\circ$).

\begin{figure}
\epsscale{0.8}
\plotone{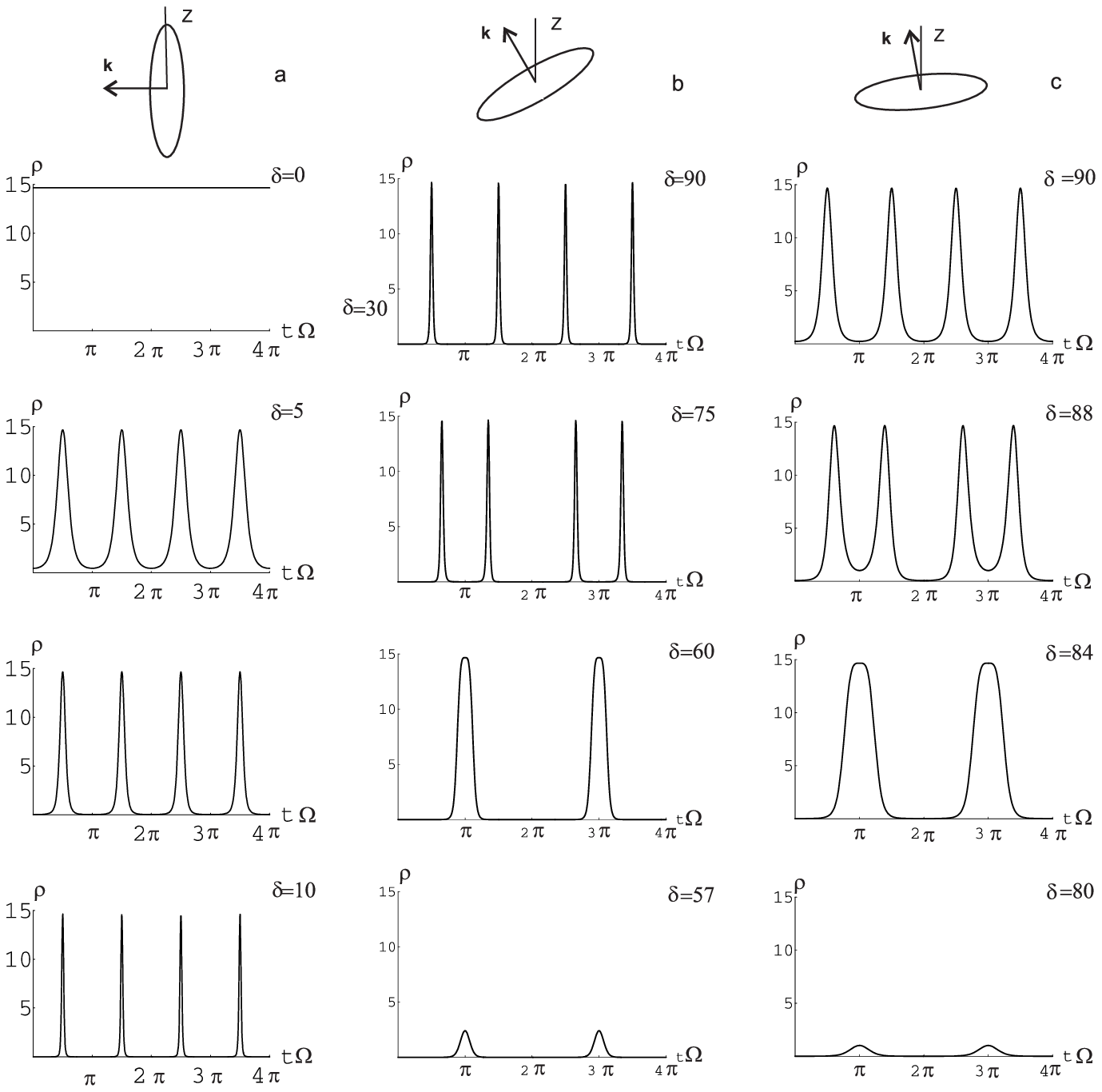}
\caption
{The radiation phase profiles for $\beta=0.99$ and angles:
a --- $\kappa=90^\circ$; b --- $\kappa=30^\circ$; c
--- $\kappa=6^\circ$}
\label{f:3}
\end{figure}

\begin{figure}
\epsscale{0.4}
\plotone{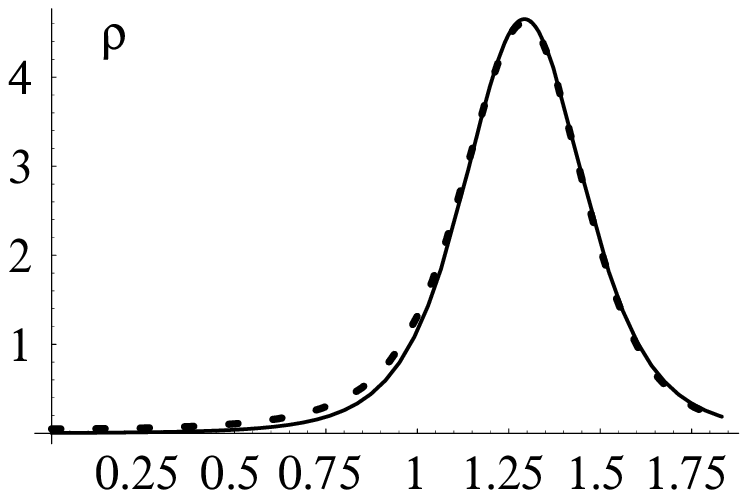}
\caption
{The profiles of radiation constructed via exact equation
(3) (dashed line) and approximation (10) (solid line)
for $\beta=0.999$ and angles $\kappa=30^\circ$, $\delta=81^\circ$}
\label{f:4}
\end{figure}

Fig.~\ref{f:3}  shows the radiation profiles of ultrarelativistic
particles ($\beta=0.99$). They are distinguished by deeper  modulation
and small pulse duration.
In Fig.~\ref{f:4} the light profiles calculated by use of approximated
equation (\ref{8}) (solid line) and exact equation (\ref{0}) with
account of (\ref{2'}) (dashed line) are shown for particles moving at the
velocity $\beta=0.999$. We see that even for
moderately relativistic particle these equation are in a good
agreement.

\section{Comparison with the observed profiles}

\begin{figure}[ht]
\epsscale{1.1}
\plottwo{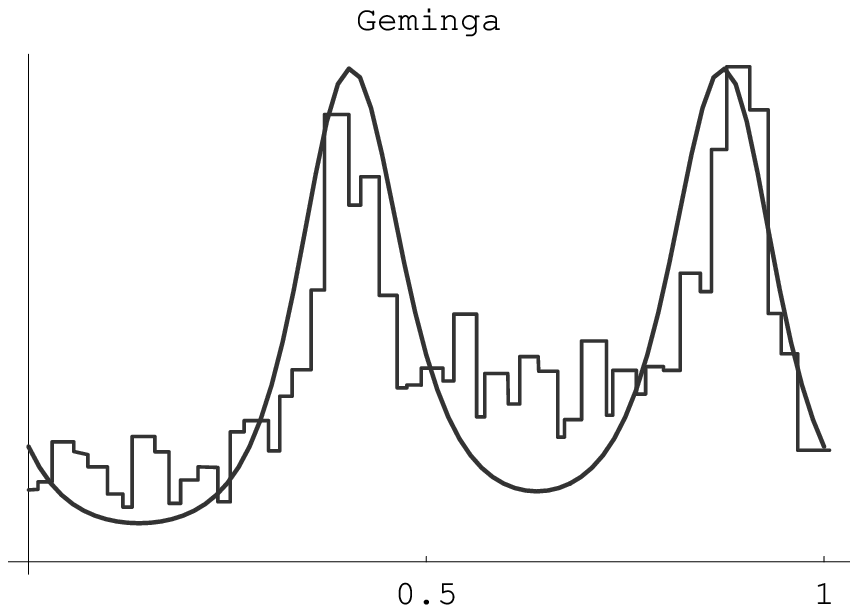}{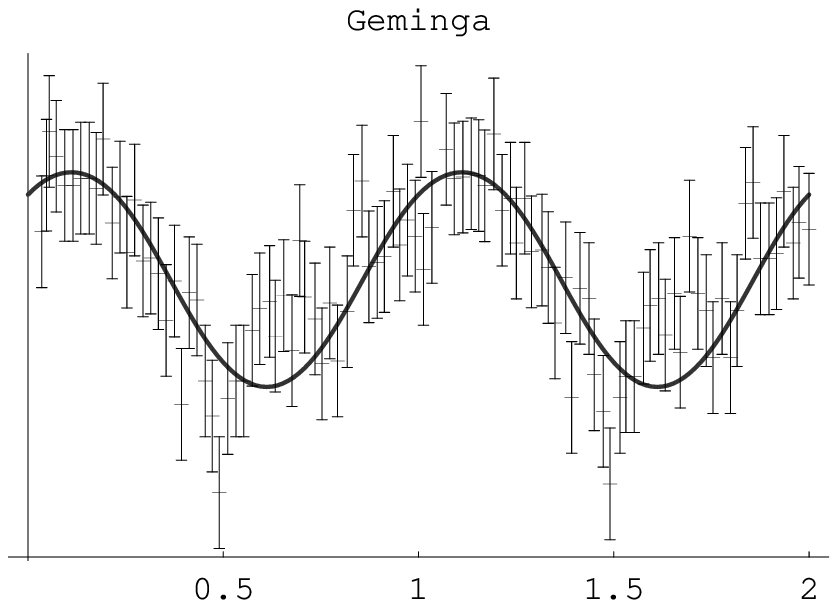}
  \caption{
Profiles  of Geminga pulsars.
Smooth lines show the model results, ghistogramms display observed
profiles. left --- $\gamma$-ray profile;
right --- X-ray profile} \label{f:6}
\end{figure}

We have considered a simple model of pulsar emission. Some of
the obtained profiles are close in shape to really observed profiles.
In this section we compare the profiles of $\gamma$-ray pulsars, the
Crab, Vela and Geminga with the profiles obtained from synchrotron
radiation indicatrix. As one can see from Figs.~\ref{f:3} and \ref{f:4}
the variation of angles $\kappa$ and $\delta$ gives a wide variety of
profiles. But it does not mean that you can produce any profile in this
way. In particular, profiles obtained by this method have a certain
relationship between the width of peaks, phase position and the level
of plato between them. Thus, not every observed profile can be built
with help of used model. But if a calculated profile fits an observed
one, it can mean that the suggested model is adequate and we obtain
some information about angles $\delta$ and $\kappa$.
\begin{figure}
\epsscale{0.8}
\plottwo{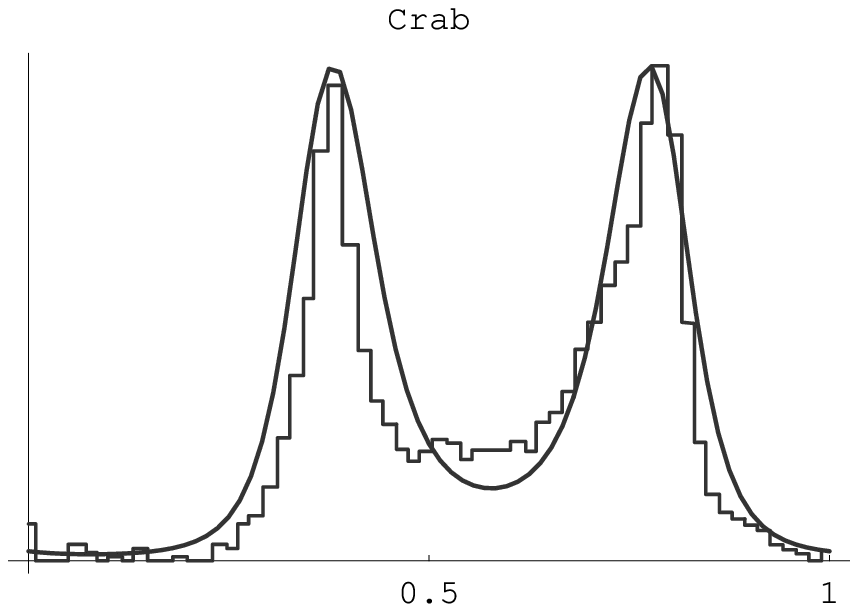}{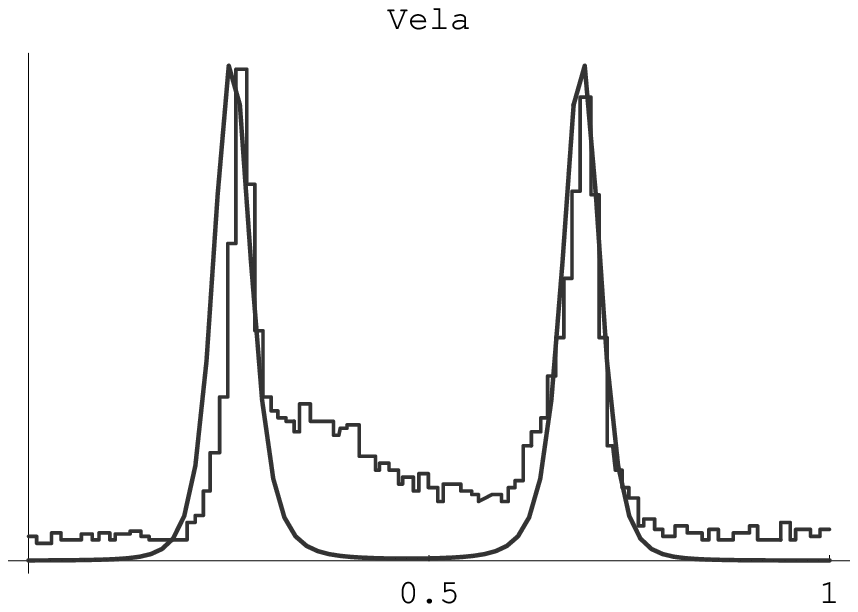}
  \caption{
Profiles  of $\gamma$-ray pulsars Crab and Vela.
Smooth lines show the model results, ghistogramms display observed
profiles.}\label{f:5}
\end{figure}
Figs~\ref{f:5} and \ref{f:6} show calculated and observed profiles for
three well known $\gamma$-ray pulsars. The energy interval for
electrons (or positron) were chosen for following reason. Most pulsars
exhibit two sharp peaks within a period if the energy of photons lies in
the region $E_\gamma\ge 1$ MeV. Knowing the critical frequency of
synchrotron radiation $\omega_\gamma\sim eH\gamma^2/mc$ we can estimate
the Lorentz-factor of particles $\gamma^2\ge mcE_\gamma/eH\hbar$. For
the magnetic field of order $H\sim 10^{12}$ Oe we obtain $\gamma\oe 10$
or for the particle velocity $\beta\ge 0.99$. Specific values of
$\beta$ for Figs \ref{5} were selected in a way to obtain the best fit
of peaks width. It is obvious that the greater is the particle energy
the narrower are the peaks.

It is believed that the pulsar in the Crab
Nebula has an inclination angle close to $\pi/2$. That is why we choose
$\kappa=85^\circ$. Then the best fit to the observed pulse profile
measured by BATSE \cite{Fishman} gives the value $\delta=15^\circ$ see
Fig. \ref{f:5}.  The next figure shows the Vela pulse profile for
energy of photons above 100 MeV detected by EGRET \cite{Fierro} along
with calculated profile for angles $\delta=10^\circ$,
$\kappa=88^\circ$ and $\beta=0.999$.  The difference occurs in a
bridge between the two pulses. It can be caused by the dispersion of
particle orbits in the pulsar magnitosphere.
 Next two figures show
Geminga profiles. The first one of Fig \ref{f:6} displays observed
$\gamma$-profile for $E_\gamma>100$ MeV \cite{Dau} along with
calculated one for $\delta=89^\circ$, $\kappa=10^\circ$ and
$\beta=0.992$. Soft X-ray profile is plotted in the second of Fig.
\ref{f:6} in the 0.53--1.5 keV band measured by ROSAT \cite{Halp} and
a profile calculated for angles $\delta=80^\circ$, $\kappa=4^\circ$ and
$\beta=0.94$. We see good agreement between the model and observed
profile though it is thought that the soft X-ray radiation is due to
thermal black body emission \cite{Halp}.

\section{Discussion}

The profiles shown on Figs \ref{f:5} and \ref{f:6} are calculated for
moderately relativistic electrons. In doing this we tried to fit the
shape of profiles.  But all the calculations were made for a single
particle.  It is evident that in a real situation radiation is emitted
by an ensemble of particles with a spread in energies and orbit
orientation.  It is evident that on averaging over this parameters the
pulse width can grow significantly. But without regard on averaging we
can obtain some information about the pulsar magnetic axis orientation
from the relative position of pulses in the radiation profile. According
to equation (\ref{3}) the maximum of radiation occurs at $t_{1,2}$
given by
\begin{equation}
\Omega
t_{1,2}=\pm\arccos(\cot\delta\cot\kappa)
\label{20}
\end{equation}
If we know the phase distance between the pulses, we will find the
relationship between angles $\delta$ and $\kappa$ by use of equation
(\ref{20}).
Here we suppose that $\gamma$-radiation of pulsars is a synchrotron
radiation of particles with their orbits orthogonal to the magnetic
axis of pulsar. The probability of detection of such pulsar in a
$\gamma$-ray band according to this model is equal to $p=\sin\kappa$.
Thus, with the most probability we can detect pulsars with its magnetic
axis orthogonal to the axis of rotation. Thereby the gap between pules
in $\gamma$-ray region should be close to 0.5 of period. That is what
we see in most cases of detected $\gamma$-ray pulsars.

The discussed model does not pretend to give entire interpretation of
the pulsar radiation. The main purpose of this paper is to show how to
find the light profile of a rotating source with an indicatrix of
radiation which is given in a rest frame. We chose the synchrotron
radiation mechanism as an widely spread example. Considered method can
be used for curvature radiation or more complicated source of radiation.

\acknowledgements

This work was supported by Ministry of General and Professional
Education of Russia, Competitive Center of Fundamental Research.


\begin{thebibliography}{99}


\bibitem[Shklovsky]{1} Shklovsky, I.S. 1977 {\it Stars}, Nauka, Moscow,
p. 384.

\bibitem[Bagrov 1965]{2}  1965 Bagrov, V.G.{\it Optika i
spektroskopija}.  V.~18. Vyp.~4. 541.

\bibitem[Bagrov et al. 1972]{3}  1972 Bagrov, V.G., Bordovitsyn, V.A.,
Kopytov, G.F.{\it Izv. Vuzov.  Fiz.} V.~15. N~6.  86.

\bibitem[Bagrov et al. 1974]{4} Bagrov, V.G., Bordovitsyn, V.A.,
Kopytov, G.F.  and Epp, V.Ya. 1974 {\it Izv.  Vuzov.  Fiz.} V.~17.
N~1. 46.

\bibitem[Bordovitsyn]{5} Bordovitsyn, V.A., ed. 1999 {\it Synchrotron
radiation and its development}.  Singapore: World Scientific.

\bibitem[Manchester and Taylor]{6}  Manchester, R.N., Taylor, J.H.
1977 {\it Pulsars}.  Freeman and Company.

\bibitem[Fishman]{Fishman}  Fishman, G. 1992 {\it GRO Newsletter}, 1, 6.

\bibitem[Fierro et al.]{Fierro}  Fierro, J.M., Michelson, P.F. and
Nolan, P.L.  1992 {\apj}, 1, 6.

\bibitem[Daugherty and Harding]{Dau} Daugherty, J.K. and Harding, A.K.
1996 {\apj} 458, 278.

\bibitem[Halpern and Ruderman]{Halp}  Halpern, J.P. and Ruderman, M.
1993 {\apj} 415, 286.

\end{thebibliography}
\end{document}